\documentclass[11pt]{article}

\usepackage{times}
\usepackage{amssymb}
\usepackage{amsmath}
\usepackage{latexsym}
\usepackage{graphics}
\usepackage{color}
\usepackage{url}

\newtheorem{theorem}{\sc Theorem}
\newtheorem{lemma}{\sc Lemma}
\newtheorem{corollary}{\sc Corollary}
\newtheorem{definition}{\sc Definition}

\newtheorem{claim}{\sc Claim}

{\scriptsize
 \bibliographystyle{abbrv}
 \bibliography{references}
}

\begin{document}

\title{Deciding k-colourability of $P_5$-free graphs in polynomial time}


\author{
Ch\'{\i}nh T. Ho\`{a}ng\thanks{Physics and Computer Science,
Wilfrid Laurier University, Canada.  Research supported by NSERC.
\texttt{ email: choang@wlu.ca}} \ \ \  \ \
Marcin~Kami\'nski\thanks{RUTCOR, Rutgers University, 640
Bartholomew Road, Piscataway, NJ 08854, USA. \texttt{email:
mkaminski@rutcor.rutgers.edu}} \ \ \ \ \
Vadim Lozin\thanks{RUTCOR, Rutgers University, 640 Bartholomew
Road, Piscataway, NJ 08854, USA. \texttt{email:
lozin@rutcor.rutgers.edu}}\ \ \ \ \
J. Sawada\thanks{Computing and Information Science, University of
Guelph, Canada. Research supported by NSERC. \texttt{ email:
sawada@cis.uoguelph.ca}} \ \ \ \ \
X. Shu\thanks{Computing and Information Science, University of
Guelph, Canada. \texttt{ email: xshu@uoguelph.ca}} \ \ \  \ \ }

\maketitle

\begin{abstract}
The problem of computing the chromatic number of a $P_5$-free graph
is known to be NP-hard. In contrast to this negative result, we show
that determining whether or not a $P_5$-free graph admits a
$k$-colouring, for each fixed number of colours $k$, can be done in
polynomial time. If such a colouring exists, our algorithm
produces~it.
\end{abstract}


\section{Introduction}
A \emph{$k$-colouring} of a graph $G$ is an assignment of $k$
colours to the vertices of $G$ so that no two adjacent vertices
receive the same colour.  The $k$-{\sc colourability} is the problem
of determining whether or not a given graph $G$ admits a
$k$-colouring. The optimization version of the problem asks to find
a $k$-colouring of $G$ with minimum $k$, called the {\it chromatic
number} of $G$ and denoted $\chi(G)$.

The $k$-{\sc colourability} is one of the central problems of
algorithmic graph theory with numerous applications \cite{DK03}. It
is also one of the most difficult problems: it is NP-complete in
general \cite{karp} and its optimization version is even hard to
approximate \cite{KLS00}. Moreover, the problem remains difficult in
many restricted graph families, for example triangle-free graphs
\cite{MP} or line graphs \cite{Degree4} (in which case it coincides
with the {\sc edge} $k$-{\sc colourability}). On the other hand,
when restricted to some other classes, such as graphs of vertex
degree at most $k$ \cite{br} or perfect graphs \cite{GLS84}, the
problem can be solved in polynomial time. Efficient polynomial-time
algorithms for finding optimal colourings are available for many
particular subclasses of perfect graphs, including chordal graphs
\cite{gavril}, weakly chordal graphs \cite{hayward}, and
comparability graphs \cite{even}.

All the aforementioned examples refer to graph classes possessing
the property that with any graph $G$ they contain all induced
subgraphs of $G$. Such classes are known in the literature under the
name of {\it hereditary} classes. Any hereditary class can be
described by a unique set of minimal graphs that do not belong to
the class, so-called {\it forbidden induced subgraphs}. A nice
survey on colouring vertices of graphs in hereditary classes can be
found in \cite{RS04}. An important line of research of this type
deals with $P_t$-free graphs, i.e., classes excluding a path on $t$
vertices $P_t$ as an induced subgraph.

Sgall and Woeginger showed in \cite{SW01} that
$5$\textsc{-colourability} is NP-complete for $P_8$-free graphs
and $4$\textsc{-coloura\-bi\-lity} is NP-complete for
$P_{12}$-free graphs. The last result was improved in
\cite{LRS06}, where the authors claim that by modifying the
reduction from \cite{SW01} $4$\textsc{-colourability} can be shown
to be NP-complete for $P_9$-free graphs. On the other hand, the
$k$-{\sc colourability} problem can be solved in polynomial time
for $P_4$-free graphs as they constitute a subclass of perfect
graphs. For $t=5,6,7$, the complexity of the problem is generally
unknown, except for the case of $3$\textsc{-colourability} of
$P_5$-free \cite{RST02,SW01} and $P_6$-free graphs \cite{RS04A}.
Known results on the $k$\textsc{-colourability} problem in classes
of $P_t$-free graphs are summarized in Table \ref{table} (under
columns 5 and 6, $\alpha$ is matrix multiplication exponent known
to satisfy $2 < \alpha < 2.376$ \cite{cop}).

  \begin{table}\label{table}
\begin{center}
\begin{tabular}{|c|c|c|c|c|c|c|c|c|c|c|c|}
\hline
$k\backslash t$&3&4&5&6&7&8&9&10&11&12&\ldots\\
\hline
3 &$O(m)$ &$O(m)$ &$O(n^\alpha)$ &$O(mn^\alpha)$&?&?&?&?&?&?&\ldots\\
4 &$O(m)$ &$O(m)$ &\bf{??}  &? &? &?      & $NP_c$ & $NP_c$& $NP_c$& $NP_c$ &\ldots\\
5 &$O(m)$ &$O(m)$ &\bf{??}  &? &? &$NP_c$ & $NP_c$ & $NP_c$& $NP_c$& $NP_c$ &\ldots\\
6 &$O(m)$ &$O(m)$ &\bf{??}  &? &? &$NP_c$ & $NP_c$ & $NP_c$& $NP_c$& $NP_c$ &\ldots\\
7 &$O(m)$ &$O(m)$ &\bf{??}  &? &? &$NP_c$ & $NP_c$ & $NP_c$& $NP_c$& $NP_c$ &\ldots\\
\ldots&\ldots&\ldots&\bf{\ldots}&\ldots&\ldots&\ldots&\ldots&\ldots&\ldots&\ldots&\ldots\\
\hline
\end{tabular}
\caption{Known complexities for $k$-colourability of $P_t$-free
graphs}
\end{center}
\end{table}

In this paper, we focus on the minimal class from Table \ref{table}
where the $k$-{\sc colourability} problem is unsolved, i.e., the
class of $P_5$-free graphs. This class is ``stubborn'' with respect
to various graph problems. For instance, $P_5$-free graphs
constitute a unique minimal class defined by a single forbidden
induced subgraph with unknown complexity of the {\sc maximum
independent set} and {\sc minimum independent dominating set}
problems. Many algorithmic problems are known to be NP-hard in the
class of $P_5$-free graphs, which includes, among others, {\sc
dominating set} \cite{Kor90} and {\sc chromatic number} \cite{kral}.
In contrast to the NP-hardness of finding the chromatic number of a
$P_5$-free graph, we show that $k$\textsc{-colourability} can be
solved in this class in polynomial time for each particular value of
$k$. In the case of a positive answer, our algorithm yields a valid
$k$-colouring. Along with the  mentioned result on
3\textsc{-colourability} of $P_5$-free graphs, our solution
generalizes several other previously studied special cases of the
problem, such as $4$\textsc{-colourability} of $(P_5, C_5)$-free
graphs \cite{LRS06} and $4$\textsc{-colourability} of $P_5$-free
graphs containing a dominating clique on four vertices \cite{wang}.
We also note the algorithm in \cite{giak} that colours a
$(P_5,\overline{P}_5)$-free graph $G$ with $\chi(G)^2$ colours.

The remainder of the paper is organized as follows.  In Section
\ref{sec:defs} we give relevant definitions, concepts, and
notations. In Section \ref{sec:algo}, we  present our recursive
polynomial time algorithm that answers the $k$-colourability
question for $P_5$-free graphs.  The difficult step in the algorithm
is detailed using two different approaches.  We conclude with a
summary of our results in Section \ref{sec:summary} along with a
list of open problems.

\section{Background and  Definitions}\label{sec:defs}

In this section we provide the necessary background and
definitions used in the rest of the paper.  For starters, we
assume that $G=(V,E)$ is a simple undirected graph where $|V| = n$
and $|E| = m$.  If $A$ is a subset of $V$, then we let $G(A)$
denote the subgraph of $G$ induced by $A$. A {\it stable set} is a
set of vertices such that there is no edge joining any two
vertices in it.

\begin{definition}  A set of vertices $A$ is said to  \emph{dominate} another set $B$, if every
vertex in $B$ is adjacent to at least one vertex in $A$.
\end{definition}

The following structural result about $P_5$-free graphs is from Bacs\'{o} and Tuza \cite{bacso}:

\begin{theorem}
\label{thm:basco}
Every connected $P_5$-free graph has either a dominating clique or
a dominating $P_3$.
\end{theorem}

\begin{definition}  Given a graph $G$,
an integer $k$ and for each vertex $v$, a list $l(v)$ of $k$
colours, the \emph{$k$-list colouring problem} asks whether or not
there is a colouring of the vertices of $G$ such that each vertex
receives a colour from its list.
\end{definition}

\begin{definition}
The \emph{restricted $k$-list colouring problem} is the $k$-list
colouring problem in which the lists $l(v)$ of colours are subsets
of $\{1,2,\ldots,k\}$.
\end{definition}

Our general approach is to take an instance of a specific colouring
problem $\Phi$ for a given graph and replace it with a polynomial
number of instances $\phi_1, \phi_2, \phi_3, \ldots $ such that the
answer to $\Phi$ is ``yes'' if and only if there is some instance
$\phi_k$ that also answers ``yes''.

For example, consider a graph with a dominating vertex $u$ where
each vertex has colour list $\{1,2,3,4,5\}$ This listing corresponds
to our initial instance $\Phi$.  Now, by considering different ways
to colour $u$, the following four instances will be equivalent to
$\Phi$:

\begin{enumerate}
\item[]$\phi_1$: \  $u=1$ and the remaining vertices have colour lists $\{2,3,4,5\}$,
\item[]$\phi_2$: \  $u=2$ and the remaining vertices have colour lists $\{1,3,4,5\}$,
\item[]$\phi_3$: \  $u=3$ and the remaining vertices have colour lists $\{1,2,4,5\}$,
\item[]$\phi_4$: \  $u=\{4,5\}$ and the remaining vertices have colour lists $\{1,2,3,4,5\}$.
\end{enumerate}

In general, if we recursively apply such an approach we would end up
with an equivalent set with an exponential number of colouring
instances.

\section{The Algorithm}
\label{sec:algo}

Let $G$ be a connected $P_5$-free graph. This section describes a
polynomial time algorithm that decides whether or not $G$ is
$k$-colourable. Our strategy is as follows. First, we find a
dominating set $D$ of $G$ which is a clique with at most $k$
vertices or a $P_3$. There are only a finite number of ways to
colour the vertices of $D$ with $k$ colours. For each of these
colourings of $D$, we recursively check if it can be extended to a
colouring of $G$. Each of these subproblems can be expressed by a
restricted list colouring problem. We now describe the algorithm in
detail.

The algorithm is outlined in 3 steps. Step 2 requires some extra
structural analysis and is presented using two different approaches in
the following subsections.


\underline{\textsf{Algorithm}}
\begin{enumerate}
\item First, we check if $G$ contains a dominating set of size at
most $k\ge3$. If no such a set is found, then $G$ is not
$k$-colourable. Otherwise, let $D$ be a dominating set in $G$, which
is either a clique with at most $k$ vertices or a $P_3$. Let the
vertices of the dominating set be $d_1, d_2, \ldots, d_t$ with $t
\leq k$. Since $D$ is a dominating set, we can partition the
remaining vertices into {\it fixed sets} $F_1, F_2, \ldots , F_r$,
$r \leq t$, such that vertices in $F_1$ are adjacent to $d_1$, and
for $i>1$, vertices in $F_i$ are adjacent to $d_i$ but not to
$\{d_1, \ldots, d_{i-1} \}$. The colour list of the vertices in the
fixed sets have size at most $k-1$ since each vertex in $D$ is
already assigned a colour. This gives rise to our original
restricted list-colouring instance $\Phi$.

\item Two vertices are \emph{dependent} if there is an edge between them
and the intersection of their colour lists is non-empty. In this
step, we remove all dependencies between each pair of fixed sets.
This process will create a set $\{\phi_1,\phi_2, \phi_3, \ldots \}$,
equivalent to $\Phi$, of a polynomial number of colouring instances.
Two different methods for performing this step are outlined in the
following subsections.

\item  For each instance $\phi_i$ from Step 2 the dependencies
between each pair of fixed sets have been removed which means that
the vertices within each fixed set can be coloured independently.
Thus, for each instance $\phi_i$ we recursively see if each fixed
set can be coloured with the corresponding restricted colour lists
(the base case is when the colour lists are a single colour). If
\emph{one} such instance provides a valid $k$-colouring then
return the colouring. Otherwise, the graph is not $k$-colourable.
\end{enumerate}
As mentioned, the difficult part is reducing the dependencies
between each pair of fixed sets (Step 2). We present two different
approaches to handle Step 2. The first is conceptually simpler
while the second includes additional structural results.

\subsection{Removing the Dependencies Between Two Fixed Sets: Method I}

Let $col(C)$ be the set of colours that appear in the lists of
vertices of a set $C$. Let $A$ and $B$ be two fixed sets. Note that
$|col(A)| \leq k-1$ and $|col(B)| \leq k-1$. We remove dependencies
between $A$ and $B$ by applying the following procedure.

\underline{\textsf{Procedure One}}
\begin{enumerate}
\item Find a $(k-1)$-colouring of $A$ (respectively, $B$)  with
stable sets $A_1, A_2, \ldots, \, A_{k-1}$ (respectively, $B_1, B_2,
\ldots,$ $ B_{k-1}$). If $A$ or $B$ cannot be $(k-1)$-coloured, then
$G$ cannot be $k$-coloured.

\item For each $i=1,2, \ldots, k-1$ and each  $j=1,2, \ldots ,
k-1$, remove dependencies between $A_i$ and $B_j$.

\end{enumerate}

Now, we describe how to remove dependencies between two
stable sets $X=A_i$ and $Y=B_j$. Let $X'$ (respectively, $Y'$) be
the set of vertices of $X$ (respectively, $Y$) that are dependent
on some vertices of $Y$ (respectively, $X$). Note that $X'$ is
non-empty if and only if $Y'$ is non-empty.

\begin{lemma}\label{dominating_vertex}
If $X' \neq \emptyset$, there exists a vertex in $X'$ that is
adjacent to all vertices in $Y'$.
\end{lemma}

\noindent \emph{Proof}. Let $x_1$ be a vertex of $X'$ with a
maximal neighborhood in $Y'$. Assume there exists a vertex $y_2
\in Y'$ that is not adjacent to $x_1$. Then, there must exist a
vertex $x_2 \in X'$ (different than $x_1$) adjacent to $y_2$. Also, by
the choice of $x_1$, there must exist a vertex $y_1 \in Y'$ that is
adjacent to $x_1$ but not $x_2$.
Since $X$ and $Y$ belong to  different
fixed sets, there exists a vertex $v$ in the dominating set such
that either $v$ is adjacent to $x_1, x_2$ but not $y_1, y_2$, or
$v$ is adjacent to $y_1, y_2$ but not $x_1, x_2$. But then $G(\{v,
x_1, x_2, y_1, y_2\})$ is an induced $P_5$; a contradiction.
$\Box$ \vspace*{.2cm}


Lemma \ref{dominating_vertex} states that as long as $X'$ and $Y'$
are non-empty, we can find a vertex $x \in X'$ that dominates
$Y'$. Now given such a vertex $x$, we can create new equivalent
colouring instances by assigning to $x$ (i) a colour from $l(x)
\cap col(Y')$ and (ii) the list $l(x) - col(Y')$.  In the former
instances the vertices in $Y'$ lose the colour assigned to $x$
from their lists i.e., $|col(Y')|$ decreases by one.  In the
latter instance, the vertex $x$ is no longer dependent on any
vertex in $Y'$ and is thus removed from $X'$.   In this case, we
recursively repeat this process until $X'$ is empty by finding a
new vertex in $X'$ that dominates $Y'$. This will result in at
most $kn$ new colouring instances where either $X'$ is empty or
$|col(Y')|$ has decreased by one from its initial state. To reduce
$|col(Y')|$ to zero, we repeatedly apply this process at most $k$
times.  Thus, we can completely remove the dependencies between
$X'$ and $Y'$ by producing at most $(kn)^k$ new equivalent
colouring instances.

{\bf Analysis.}  To remove the dependencies between each $A_i$ and
$B_j$ requires $(kn)^k$ new equivalent instances.  Thus, to remove
the dependencies between each pair of fixed sets (Step 2 of
\textsf{Procedure One})  requires $(kn)^{k^3}$ new equivalent
instances.  Since there are $k$ fixed sets, there are less than
$k^2$ pairs of fixed sets.  Thus, to remove dependencies between
each pair of fixed sets (given the stable sets for each fixed set)
requires $(kn)^{k^5}$ equivalent instances.  To find the stable sets
for each fixed set requires a single recursive $k{-}1$ colouring on
the graph $G$ with the initial dominating set combined with the
edges between the fixed sets removed.

Now, let $T(k)$ denote the number of subproblems produced by the
\textsf{Algorithm} where $k$ is the number of colours used on a
graph with $n$ vertices.  From the previous analysis we arrive at
the following recurrence where $T(1) = 1$:
\[T(k) = (kn)^5 T(k-1) + T(k-1). \]
A proof by induction shows $T(k) = O((kn)^{k^6})$, implying our algorithm
runs in polynomial time.

\subsection{Removing the Dependencies Between Two Fixed Sets: Method II}

For our second method for removing the dependencies between a pair
of fixed sets, it will be convenient to associate a fixed set $F_i$
to the colours in its lists. For this purpose,  let $S_{list}$
denote a fixed set of vertices with colour list given by $list$. We
partition each such fixed set into {\bf dynamic sets} $P_i$ that
each represents a unique subset of the colours in $list$. For
example: $S_{123} = P_{123} \cup P_{12} \cup P_{13} \cup P_{23} \cup
P_1 \cup P_2 \cup P_3$. Initially, $S_{123} = P_{123}$ and the
remaining sets in the partition are empty. However, as we start
removing dependencies, these sets will dynamically change. For
example, if a vertex $u$ is initially in $P_{123}$ and one of its
neighbors gets coloured 2, then $u$ will be removed from $P_{123}$
and added to $P_{13}$.

Recall that our goal is to remove the dependencies between two
fixed sets $S_p$ and $S_q$.  To do this, we remove the
dependencies between each pair  ($P,Q$) where $P$ is a dynamic
subset of $S_p$ and $Q$ is a dynamic subset of $S_q$.  By visiting
these pairs in order from largest to smallest with respect to
$|col(P)|$ and then $|col(Q)|$, we ensure that we only need to
consider each pair once.
Applying this approach, the crux of the reduction process is to
remove the dependencies between a pair $(P,Q)$ by creating at most a
polynomial number of equivalent colourings.

\begin{figure}
  \begin{center}
  \resizebox{!}{2.2in}{\includegraphics{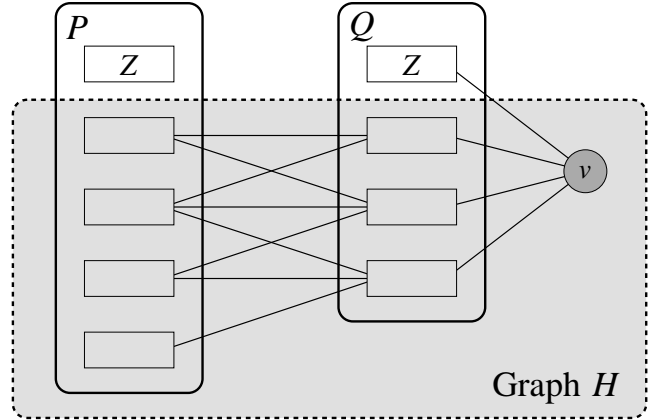}}
  \end{center}
  \caption{Illustration of the graph $H$ from two dynamic sets}
  \label{fig:reduction}
\end{figure}

Now, observe that there exists a vertex $v$ from the dominating
set found in Step 1 of the algorithm that dominates every vertex
in one set, but is not adjacent to any vertex in the other.  This
is because $P$ and $Q$ are subsets of different fixed sets.
Without loss of generality assume that $v$ dominates $Q$. Now,
consider the (connected) components of $G(P)$ and $G(Q)$.  If a
component $Z$ in $G(P)$ is not adjacent to any vertex in $Q$ then
the vertices in $Z$ have no dependencies with $Q$.  The same
applies for such components in $Q$. Since these components have no
dependencies, we focus on the induced subgraph $H = G(P \cup Q
\cup \{v\})$ with these components removed. This graph is
illustrated in Figure \ref{fig:reduction} where the small
rectangles represent the components in $G(P)$ and $G(Q)$
respectively.   It is easy to observe that $H$ is connected (if
not, then there are components $H_1, H_2$ of $H$, each of which
contains a vertex in $P$ and a vertex in $Q$; it follows there are
edges $(a,b)$ of $H_1$ and $(c,d)$ of $H_2$ such that $a,b,v,d,c$
induce a $P_5$).

\begin{theorem}
\label{thm:main} Let $H$ be a connected $P_5$-free graph partitioned
into three sets $P$, $Q$ and $\{v\}$ where $v$ is adjacent to every
vertex in $Q$ but not adjacent to any vertex in $P$.
Then there exists at most one component in $G(P)$ that contains two
vertices $a$ and $b$ such that $a$ is adjacent to some component
$Y_1 \in G(Q)$ but not adjacent to another component $Y_2 \in G(Q)$
while  $b$ is adjacent to $Y_2$ but not $Y_1$.
\end{theorem}

{\sc Proof:}  The proof is by contradiction.  Suppose that there are
two unique components $X_1,X_2 \in G(P)$ with $a,b \in X_1$ and $c,d
\in X_2$ and components $Y_1 \ne Y_2$ and $Y_3 \ne Y_4$ from $G(Q)$
such that:
\begin{itemize}
\item $a$ is adjacent to $Y_1$ but not adjacent to $Y_2$,
\item $b$ is adjacent to $Y_2$ but not adjacent to $Y_1$,
\item $c$ is adjacent to $Y_3$ but not adjacent to $Y_4$,
\item $d$ is adjacent to $Y_4$ but not adjacent to $Y_3$.
\end{itemize}

\begin{figure}
  \begin{center}
  \resizebox{!}{2.0in}{\includegraphics{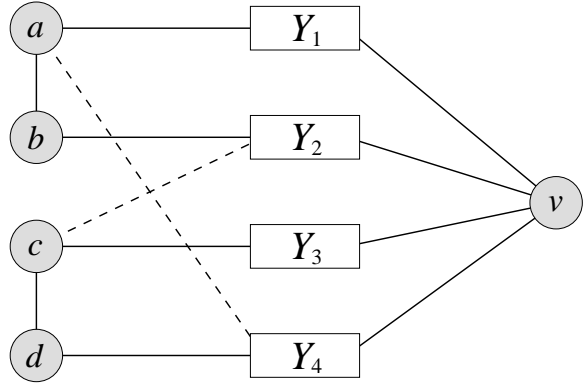}}
  \end{center}
  \caption{Illustration for proof of Theorem \ref{thm:main}}
  \label{fig:proof}
\end{figure}

Let $y_i$ denote an arbitrary vertex from the component $Y_i$.
Since $H$ is $P_5$-free, there must be edges $(a,b)$ and $(c,d)$,
otherwise $a,y_1,v,y_2,b$ and $c,y_3,v,y_4,d$ would be $P_5$s. An
illustration of these vertices and components is given in Figure
\ref{fig:proof} - the solid lines.

Now, if $Y_2 =Y_3$, then there exists a $P_5 = a,b,y_2,c,d$. Thus,
$Y_2$ and $Y_3$ must be unique components, and $Y_1, Y_4$ must be
different as well for the same reason. Similarly $Y_2 \ne Y_4$. Now
since $b,y_2,v,y_3,c$ cannot be a $P_5$, either $b$ is adjacent to
$Y_3$ or $c$ must be adjacent to $Y_2$. Without loss of generality,
suppose the latter. Now $a,b,y_2,v,y_4$ implies that either $a$ or
$b$ is adjacent to $Y_4$. If the latter, then $a,b,y_4,d,c$ would be
a $P_5$ which implies that $a$ must be adjacent to $Y_4$ anyway.
Thus, we end up with a $P_5 = a,y_4,v,y_2,c$ which is a
contradiction to the graph being $P_5$-free. \hfill $\Box$

From Theorem \ref{thm:main}, there is at most one component $X$ in $G(P)$ that contains two vertices $a$ and $b$
such that $a$ is adjacent to some component $Y_1 \in G(Q)$ but not
adjacent to another component $Y_2 \in G(Q)$ while  $b$ is adjacent
to $Y_2$ but not $Y_1$. If such a component exists, then we can
remove the vertices in $X$ from $P$ by applying the following general method for removing a component $C$
from a dynamic set $D$.

\noindent \underline{\textsf{Procedure RemoveComponent}}
\begin{itemize}
\item[] Since $C$ is $P_5$-free, it has a dominating clique or
$P_3$ (Theorem \ref{thm:basco}). If this dominating set $D$ can be
coloured with the list $col(D)$, we consider all such colourings
(otherwise we report there is no valid colouring for the given
instance). For each case the colouring will remove all vertices in
the component from $D$ to other dynamic sets represented by smaller
subsets of available colours. Observe that since $k$ is fixed, the
number of such colourings is constant.
\end{itemize}

If there are still dependencies between $P$ and $Q$, then we make the following claim
(observing that the graph $H$ dynamically changes as $P$ and $Q$ change):

\begin{claim} There exists a vertex $x \in P$ that is adjacent to all components in $H(Q)$.
Moreover, $x$ dominates all components of $H(Q)$ except at most one.
\end{claim}
{\sc Proof:} Let $x \in P$ be adjacent to a maximal number of
components in $H(Q)$.  If it is not adjacent to all components, then
there must exist another vertex $x' \in P$ and components $Y_1,Y_2
\in Q$ such that $x$ is adjacent to $Y_1$ but not $Y_2$ and $x'$
is adjacent to $Y_2$ but not $Y_1$.  This implies that there is a
$P_5 = x,y_1,v,y_2,x'$  where $y_1 \in Y_1$ and $y_2 \in Y_2$
unless $x$ and $x'$ are adjacent.  However by Theorem
\ref{thm:main}, they cannot belong to the same component in $H(P)$
since such a component would already have been removed - a
contradiction.

Now, suppose that there are two components $Y_1$ and $Y_2$ in $H(Q)$ that $x$ does \emph{not} dominate.
Then there exists edges $(y_1, y_1') \in Y_1$ and $(y_2,y_2') \in Y_2$ such that $x$ is adjacent to
$y_1$ and $y_2$, but not $y_1'$ nor $y_2'$.  This however, implies the $P_5 = y_1', y_1,x,y_2,y_2'$ -
a contradiction.
\hfill $\Box$

Now we identify such an $x$ outlined in this claim and create
equivalent new colouring instances by assigning $x$ with each colour
from $col(P) \cap col(Q)$ and then with the list $col(P)-col(Q)$. If
$x$ is assigned a colour from $col(P) \cap col(Q)$, then all but at
most one component will be removed from  $H(Q)$. If one component
remains, then we can remove it from $Q$ by applying
\textsf{Procedure RemoveComponent}.  In the latter case, where $x$
is assigned the colour list  $col(P) - col(Q)$, $x$ will be removed
from $P$.  If there are still dependencies between $P$ and $Q$, we
repeat this step by finding another vertex $x$. In the worst case we
have to repeat this step at most $|P|$ times. Therefore, the process
for removing the dependencies between two dynamic sets creates at
most $O(n)$ new equivalent colouring instances.

{\bf Analysis.}  We have just shown that we require at most $O(n)$
new equivalent colouring instances to remove the dependencies
between two dynamic sets.  Since each fixed set contains at most
$O(2^{k-1})$ dynamic sets, there are $O(2^{2(k-1)})$ pairs of
dynamic sets to consider between each pair of fixed sets.  Thus,
removing the dependencies between two fixed sets produces
$O(n^{2^{2(k-1)}})$ subproblems.  Since there at most $k^2$ pairs of
fixed sets, this means that to remove the dependencies between all
fixed sets creates $O(n^{k^2\cdot 2^{2(k-1)}})$ subproblems.

As with the previous method, let $T(k)$ denote the number of
subproblems produced by the \textsf{Algorithm} where $k$ is the
number of colours used on a graph with $n$ vertices.   From the
previous analysis we arrive at the following recurrence where $T(1)
= 1$:
\[ T(k) \leq cn^{k^2 \cdot 2^{2(k-1)}} T(k-1).\]
A proof by induction proves that $T(k)= O( n^{k^3 \cdot
4^{(k-1)}})$, implying our algorithm runs in polynomial time.

%
%
%
%
%


\begin{theorem}\label{thm:main2}
The restricted $k$-list colouring problem for $P_5$-free graphs, for
a fixed integer $k$, can be solved in polynomial time.
\end{theorem}
\begin{corollary}
Determining whether or not a $P_5$-free graph can be coloured with
$k$-colours, for a fixed integer $k$, can be decided in polynomial
time.
\end{corollary}

\section{Summary}
\label{sec:summary}

The algorithm presented in this paper brings us one step closer to
completely answering the question of when there exists a
polynomial time algorithm for the $k$\textsc{-colourability}
problem for $P_t$-free graphs, given fixed $k$ and $t$.  In
particular, we now know that there exists a polynomial time
algorithm when $t=5$ for any fixed value of $k$.

Continuing with this vein of research, the
following open problems are perhaps the next interesting avenues for
future research:
\begin{itemize}
\item Does there exist a polynomial time algorithm determine whether or not
a $P_7$-free graph can 3-coloured.
\item Does there exist a polynomial time algorithm determine whether or not
a $P_6$-free graph can 4-coloured.
\item Is the problem of $k$-colouring a $P_7$-free graph NP-complete.
\end{itemize}
Two other related open problems are to determine the complexities
of the {\sc maximum independent set} and {\sc minimum independent
dominating set} problems on $P_5$-free graphs.


\end{document}